\UseRawInputEncoding
\documentclass[12pt, a4paper]{article}
\title{Implications of New Quantum Spin Perspective In Quantum Gravity}
\author{Rakshit P. Vyas$ ^1 $, Mihir J. Joshi$ ^2 $}
\date {%
$^1$ \textit{Department of Physics, Saurashtra University, Rajkot, India} \\%
$^2$ \textit{Department of Physics, Saurashtra University,  Rajkot, India} \\%
\today %
}

\usepackage{blindtext}
\usepackage{geometry}
\usepackage{amsmath}
\usepackage{amssymb}
\usepackage{amsfonts}
\usepackage{graphicx}
\usepackage{stabular}

\begin{document}
\maketitle
\begin{abstract}

Consequences of new quantum spin perspective in quantum gravity are far-reaching. Results of this novel perspective in loop quantum gravity, i.e., the modification of the equation of geometrical operators such as the area and the volume operator are known. Using newly proposed formula from this perspective, the magnitude of fundamental constants such as the reduced Planck constant \(\hbar\), the gravitational constant \(G\), the speed of light \(c\), the Boltzmann constant \(k_{\beta}\), the fine structure constant \(\alpha\), can be validated. With the aid of this perspective, we find new formulas for the fundamental Planckian quantities and the derived Planckian quantities. We also propose novel formulas for the Planck star such as the size, the curvature, the surface area and the size of black hole (for the Planck star)  without modifying its significance. The relation of the quantum spin with the Planck temperature \(T_{P}\) \((T_{p} \propto n^{2})\), the Planck mass \(m_{P}\) \((m_{P} \propto n^{2})\), the Planck length \(l_{P}\) \((l_{P} \propto n)\) are also proposed using this novel perspective.   
  

\end{abstract}

\maketitle

\section*{Introduction}

The quantum spin is the crucial intrinsic concept or property in quantum physics. Penrose [1, 2] gave the notion of spin network in which the quantum spin is used to explain the space-time in discrete way. The spin network is the building block of loop quantum gravity (LQG) [3 -10]. LQG quantizes geometrical observables such as area and volume [11-13].  Earlier, we [14] proposed a new perspective of quantum spin and applied it to the quantum geometry using the spin network. Using this perspective, the reduced Planck constant \(\hbar\) can be defined in novel way. 

In thermodynamics, the temperature is expressed as average kinetic energy of system of particles [15] i.e., 

\begin{eqnarray}
\frac{1}{2}mv^{2} = \frac{3}{2}k_{\beta}T
\end{eqnarray}

Where, \(k_{\beta}\) is the Boltzmann constant. If the equi-partition theorem is taken into consideration; then, one the kinetic energy i.e., \(\frac{1}{2}k_{\beta}T\) for each degrees of freedom (\(1D\)).  By multiplying the numerator and denominator of left side of the equation (1) by mass \(m\) and, thereafter, both sides by \(r^{2}\); one obtains

\begin{eqnarray}
\frac{p^{2}r^{2}}{2m} = \frac{k_{\beta}T r^{2}}{2}
\end{eqnarray}

In classical physics,  the angular momentum (the scalar form) is written as \(l = rp\); hence, the equation takes the form,  

\begin{eqnarray}
\therefore l^{2} = k_{\beta}T r^{2} m
\end{eqnarray}

Till now, domain of classical physics is considered. If Bohr's hypothesis [16] of the angular momentum quantization is taken into consideration; then, one enters into quantum domain. i.e., 

\begin{eqnarray}
l^{2} = n^{2} \hbar^{2} = k_{\beta}T r^{2} m
\end{eqnarray}

Where \(\hbar =\frac{h}{2\pi}\), is the reduced Planck constant. If \(r\), \(T\) and \(m\) are replaced by Planckian quantities i.e.,  \(r = l_{P}\), \(T = T_{P}\) and \(m = m_{P}\) respectively; one enters into the domain of the quantum gravity [17-20]. Thus, 

\begin{eqnarray}
l^{2} = n^{2} \hbar^{2} = k_{\beta}T_{P} l_{P}^{2} m_{P}
\end{eqnarray}

In the theory of spin network, the total angular momentum \(J\) is more essential than the angular momentum \(l\); since, \(z\)-direction is unknown priory. Hence, the value of \(j\) plays an essential role than the value of \(m\). So, the total angular momentum \(J\) is applied to the equation (5). In the theory of spin network, the quantum spin (\(J\)) is expressed as twice of the actual quantum spin \(\frac{n\hbar}{2}\) [1-2].

\begin{eqnarray}
J = 2 \times \left(\frac{n \hbar}{2}\right)
\end{eqnarray}

Where, \(J\) is the total angular momentum and \(n\)  is integer.  Hence, from the equation (5) and (6), one gets,

\begin{eqnarray}
J^{2} = 2^{2} \times \left(\frac{n^{2} \hbar^{2}}{2^{2}}\right) = k_{\beta}T_{P} l_{P}^{2} m_{P}
\end{eqnarray}

The equation (7) is validated in [14]. In equation (7), if, \(n = 1\) is taken into consideration; then, one gets the novel definition of the reduced planck constant \(\hbar\), i.e., 

\begin{eqnarray}
\hbar = \sqrt{k_{\beta}T_{P} l_{P}^{2} m_{P}} = 1.05375 \times 10^{-34} J \cdot s \approx 1.05457 \times 10^{-34} J \cdot s
\end{eqnarray}

From equation (8), one can say that the obtained value of the \(\hbar\) is approximately equal to the actual value of \(\hbar\). In the physics, the maximum permitted value for the quantum spin is from \(\frac{1}{2}\) to \(2\). Accordingly, the value of the integer \(n\) (in equation (7)) is \(n = 1, 2, 3, 4\). At quantum gravity scale, the consequences of this novel quantum spin perspective can be found in the following way.

\section*{Universal constants from quantum spin}

Using this novel perspective, we propose new formulas for fundamental physical constants such as the gravitational constant \(G\), the speed of light \(c\), the Boltzmann constant \(k_{\beta}\) and also validate the value of these constants. For this purpose, \(n=1\) is taken into equation (7).

From equation (7), the value of the Boltzmann constant \(k_{\beta}\)  can be found as, 

\begin{eqnarray}
k_{\beta} = \frac{\hbar^{2}}{T_{P} l_{P}^{2} m_{P}} \approx 1.3806 \times 10^{-23} J/K (\because n=1)
\end{eqnarray}

From this novel formula, the obtained value of the Boltzmann constant \(k_{\beta}\) is approximately equal to the actual value. One can also calculate the value of the universal gravitation constant \(G\),  and the speed of light \(c\) by comparing the derived formula of the Planck temperature \(T_{P}\) from the quantum spin (from eqn. (7)) with the old formula of the Planck temperature [14].

The actual formula of the Planck temperature is [17-20] 

\begin{eqnarray}
T_{P} = \sqrt{\frac{\hbar c^{5}}{G k_{\beta}^{2}}}
\end{eqnarray}

And the derived formula of the Planck temperature from equation (7) is written as 

\begin{eqnarray}
T_{P} = \frac{2^{2} \times \left(\frac{n^{2} \hbar^{2}}{2^{2}}\right)}{k_{\beta} l_{P}^{2} m_{P}}
\end{eqnarray} 
 
By comparing the equation (10) and (11), one gets,

\begin{eqnarray}
\sqrt{\frac{\hbar c^{5}}{G k_{\beta}^{2}}} = \frac{2^{2} \times \left(\frac{n^{2} \hbar^{2}}{2^{2}}\right)}{k_{\beta} l_{P}^{2} m_{P}}
\end{eqnarray}  

One can derive the novel formula for \(G\),  and \(c\) from equation (12); i.e.,

\begin{eqnarray}
G = \frac{c^{5} m_{P}^{2}l_{P}^{4}}{\hbar^{3}} \approx 6.70 \times 10^{-11} Nm^{2}/kg^{2}
\end{eqnarray}

\begin{eqnarray}
c  = \sqrt[5]{\frac{\hbar^{3} G}{l_{P}^{4}m_{P}^{2}}} \approx 2.9999 \times 10^{8} m/s
\end{eqnarray}

One can also derive the novel formula for the fine structure constant \(\alpha\) using equation (8).  The actual formula of the fine structure constant \(\alpha\) is written as,

\begin{eqnarray}
\alpha = \frac{1}{4 \pi \varepsilon_{0}}\frac{e^{2}}{\hbar c} 
\end{eqnarray}

The new formula of the fine structure constant \(\alpha\) from this perspective can also be found as, 

\begin{eqnarray}
\alpha = \frac{1}{4 \pi \varepsilon_{0}}\frac{e^{2}}{\sqrt{k_{\beta}T_{P} l_{P}^{2} m_{P}} c} = 0.00730597 \approx 0.00729735 
\end{eqnarray} 

These values are almost equal to actual values of these constants [17-20]. This comparison suggests a good agreement with previous formalism. In the tabular form, the old formula are compared with the new formula of the fundamental constants; i.e.,

\begin{center}
\begin{tabular}{{|p{0.03\textwidth} | p{0.2\textwidth} | p{0.12\textwidth}|p{0.2\textwidth} |}}
\hline
Sr. No. & Fundamental constants & Old formula & New formula from quantum spin \\
\hline\hline
1 & Reduced Planck constant & \(\hbar = \frac{E}{\omega}\) & \(\hbar = \sqrt{k_{\beta}T_{P} l_{P}^{2} m_{P}}\) \\
\hline
2 & Boltzmann constant & \(k_{\beta} = \frac{R}{N_{A}}\) & \(k_{\beta} = \frac{\hbar^{2}}{T_{P} l_{P}^{2} m_{P}}\) \\
\hline
3 & Gravitational constant & \(G = \frac{F r^{2}}{m^{2}}\) & \(G = \frac{c^{5} m_{P}^{2}l_{P}^{4}}{\hbar^{3}}\) \\
\hline
4 & Speed of light & \(c = \frac{1}{\mu_{0}\varepsilon_{0}}\) & \(c = \sqrt[5]{\frac{\hbar^{3} G}{l_{P}^{4}m_{P}^{2}}}\) \\
\hline
5 & fine structure constant & \(\frac{1}{4 \pi \varepsilon_{0}}\frac{e^{2}}{\hbar c}\) & \(\frac{1}{4 \pi \varepsilon_{0}}\frac{e^{2}}{\sqrt{k_{\beta}T_{P} l_{P}^{2} m_{P}} c}\) \\
\hline
\end{tabular}
\end{center}
\begin{center}
Table 1: Comparison between the old formula and new formula of fundamental constant.
\end{center}

Further, we also propose new formulas for the fundamental  Planckian quantities and the derived Planckian quantities. 

\section*{The fundamental Planck units from quantum spin}

One can also derive novel formulas from fundamental Planckian quantities from equation (7) and (8). For instance, the actual formula of the Planck length is written as [17-20], 

\begin{eqnarray}
l_{P} = \sqrt{\frac{\hbar G}{c^{3}}} = 1.616 \times 10^{-35} m
\end{eqnarray}

From equation (7), the derived novel formula for the Planck length is,

\begin{eqnarray}
l_{P} = \sqrt{\frac{\hbar^{2}}{k_{\beta} T_{P} m_{P}}} = 1.618 \times 10^{-35} m
\end{eqnarray}

Similar to the Planck length, the actual formula of the Planck mass is written as [17-20],

\begin{eqnarray}
m_{P} = \sqrt{\frac{\hbar c}{G}} = 2.176 \times 10^{-8} kg
\end{eqnarray}

From equation (7), the derived novel formula for the Planck mass is,

\begin{eqnarray}
m_{P} = \frac{\hbar^{2}}{k_{\beta} l_{P}^{2} T_{P}} = 2.181 \times 10^{-8} kg
\end{eqnarray}

the actual formula of the Planck temperature is written as [17-20],

\begin{eqnarray}
T_{P} = \sqrt{\frac{\hbar c^{5}}{G k_{\beta}^{2}}} = 1.416 \times 10^{32} K
\end{eqnarray}

From equation (7), the derived novel formula for the Planck temperature is,

\begin{eqnarray}
T_{P} = \frac{\hbar^{2}}{k_{\beta} l_{P}^{2} m_{P}} = 1.419 \times 10^{32} K
\end{eqnarray}

the actual formula of the Planck time is written as [17-20],

\begin{eqnarray}
t_{P} = \sqrt{\frac{\hbar G}{c^{5}}} = 5.391 \times 10^{-44} s
\end{eqnarray}

From equation (8), the derived novel formula for the Planck time is,

\begin{eqnarray}
t_{P} = \sqrt{\frac{\left(k_{\beta}T_{P} l_{P}^{2} m_{P}\right)^{\frac{1}{2}} G}{c^{5}}} =5.378 \times 10^{-44} s
\end{eqnarray}

  One can compare the old formulas fundamental Planck quantities [17-20] with the newly derived formula from new quantum spin perspective in the tabular form. This table suggests a good agreement of the numerical values between the old fundamental Planck quantities and new fundamental Planck quantities.

\begin{center}
\begin{tabular}{{|p{0.03\textwidth} | p{0.15\textwidth} | p{0.21\textwidth}|p{0.3\textwidth} |}}
\hline
\textbf{Sr. No.} & \textbf{Planck units} & \textbf{Old formula} & \textbf{New formula from quantum spin} \\
\hline\hline
1 & Planck length & \(l_{P} = \sqrt{\frac{\hbar G}{c^{3}}} =  1.616 \times 10^{-35}\) \(m\) & \(l_{P} = \sqrt{\frac{\hbar^{2}}{k_{\beta} T_{P} m_{P}}} = 1.618 \times 10^{-35}\) \(m\) \\
\hline
2 & Planck mass & \(m_{P} = \sqrt{\frac{\hbar c}{G}}= 2.176 \times 10^{-8}\) \(kg\) & \(m_{P} = \frac{\hbar^{2}}{k_{\beta} l_{P}^{2} T_{P}}= 2.181 \times 10^{-8}\) \(kg\) \\
\hline
3 & Planck Temperature & \(T_{P} = \sqrt{\frac{\hbar c^{5}}{G k_{\beta}^{2}}} = 1.416 \times 10^{32}\) \(K\) & \(T_{P} = \frac{\hbar^{2}}{k_{\beta} l_{P}^{2} m_{P}} =  1.419 \times 10^{32}\) \(K\) \\
\hline
4 & Planck time & \(t_{P} = \sqrt{\frac{\hbar G}{c^{5}}}= 5.391 \times 10^{-44}\) \(s\) & \(t_{P} = \sqrt{\frac{\left(k_{\beta}T_{P} l_{P}^{2} m_{P}\right)^{\frac{1}{2}} G}{c^{5}}} = 5.378 \times 10^{-44}\) \(s\) \\
\hline
\end{tabular}
\end{center}
\begin{center}
Table 2: Fundamental Planck scale physical quantities.
\end{center}

\section*{The derived Planck units from quantum spin}

The new set of formulas for the derived Planck quantities can also be derived using novel quantum spin perspective. 

The actual formula of the Planck charge \(Q_{P}\) is written as [17-20], 

\begin{eqnarray}
Q_{p} = \sqrt{4 \pi \varepsilon_{0} \hbar c} = 1.8755 \times 10^{-18} C
\end{eqnarray}

From equation (8), one gets new formula for the Planck charge; i.e., 

\begin{eqnarray}
Q_{P} = \left(4 \pi \varepsilon_{0} \sqrt{k_{\beta}T_{P} l_{P}^{2} m_{P}}      c\right)^{\frac{1}{2}} = 1.8745 \times 10^{-18} C
\end{eqnarray}

The actual formula of the Planck momentum \(P_{P}\) is written as [17-20],

\begin{eqnarray}
P_{P} = m_{P}c = 6.5249 kg\cdot m/s
\end{eqnarray}

From equation (7), one gets new formula for the Planck momentum; i.e.,

\begin{eqnarray}
P_{P} = \frac{\hbar^{2} c}{k_{\beta} T_{P} l_{P}^{2}} = 6.5310 kg\cdot m/s
\end{eqnarray}

The actual formula of the Planck energy \(E_{P}\) is written as [17-20],

\begin{eqnarray}
E_{P} = m_{P}c^{2} = 1.9561 \times 10^{9} J
\end{eqnarray}

From equation (20), one gets new formula for the Planck energy; i.e.,

\begin{eqnarray}
E_{P} = \frac{\hbar^{2} c^{2}}{k_{\beta} T_{P}l_{P}^{2}} = 1.9593 \times 10^{9} J
\end{eqnarray}

The actual formula of the Planck density \(\rho_{P}\) is written as [17-20],

\begin{eqnarray}
\rho = \frac{m_{P}}{l_{P}^{3}} = 5.1550 \times 10^{96} kg/ m^{3}
\end{eqnarray}

From equation (18), one gets new formula for the Planck density; i.e.,

\begin{eqnarray}
\rho_{P} = m_{P}\cdot \left(\frac{k_{\beta} T_{P}m_{P}}{\hbar^{2}}\right)^{\frac{3}{2}} = 5.1527 \times 10^{96} kg/m^{3}
\end{eqnarray}

The actual formula of the Planck acceleration \(a_{P}\) is written as [17-20],

\begin{eqnarray}
a_{P} = \frac{c}{t_{P}} = 5.5608 \times 10^{51}(m/s^{2} 
\end{eqnarray}

From equation (24), one gets new formula for the Planck acceleration; i.e.,

\begin{eqnarray}
a_{P} = c \cdot \left(\frac{c^{5}}{\sqrt{k_{\beta}T_{P} l_{P}^{2} m_{P}} G}\right)^{\frac{1}{2}} = 5.5783 \times 10^{51} m/s^{2}
\end{eqnarray}

The actual formula of the Planck force \(F_{P}\) is written as [17-20],

\begin{eqnarray}
F_{P} = \frac{E_{P}}{l_{P}} = 1.2103 \times 10^{44} N
\end{eqnarray}

From equation (18), one gets new formula for the Planck force; i.e.,

\begin{eqnarray}
F_{P} = E_{P}\cdot \sqrt{\frac{k_{\beta}m_{P}T_{P}}{\hbar^{2}}} = 1.2109 \times 10^{44} N
\end{eqnarray}

The actual formula of the Planck frequency \(f_{P}\) is written as [17-20],

\begin{eqnarray}
f_{P} = \frac{c}{l_{P}} = 1.8549 \times 10^{43} Hz
\end{eqnarray}

From equation (18), one gets new formula for the Planck frequency; i.e.,

\begin{eqnarray}
f_{P} = \frac{c \sqrt{k_{\beta} T_{P}m_{P}}}{\hbar} = 1.8560 \times 10^{43}  Hz
\end{eqnarray}

Here, In the tabular form, the derived Planck quantities [17-20] are compared with the newly derived formulas from new quantum spin perspective. This table suggests a good agreement of the numerical values between the old derived Planck quantities and new derived Planck quantities.

\begin{center}
\begin{tabular}{{|p{0.03\textwidth} | p{0.16\textwidth} | p{0.21\textwidth}|p{0.3\textwidth}|}}
\hline
\textbf{Sr. No.} & \textbf{Derived Planck units} & \textbf{Old formula} & \textbf{New formula from quantum spin}  \\
\hline\hline

1 & Planck density & \(\rho = \frac{m_{P}}{l_{P}^{3}} = 5.1550 \times 10^{96}\) \(kg/ m^{3}\) & \(\rho_{P} = m_{P}\cdot \left(\frac{k_{\beta} T_{P}m_{P}}{\hbar^{2}}\right)^{\frac{3}{2}} = 5.1527 \times 10^{96}\) \(kg/m^{3}\)  \\
\hline

2 & Planck acceleration  & \(a_{P} = \frac{c}{t_{P}} = 5.5608 \times 10^{51}\) \(m/s^{2}\) & \(a_{P} = c \cdot \left(\frac{c^{5}}{\sqrt{k_{\beta}T_{P} l_{P}^{2} m_{P}} G}\right)^{\frac{1}{2}} = 5.5783 \times 10^{51}\) \(m/s^{2}\)  \\
\hline

3 & Planck momentum & \(P_{P} = m_{P}c = 6.5249\) \(kg\cdot m/s\) & \(P_{P} = \frac{\hbar^{2} c}{k_{\beta} T_{P} l_{P}^{2}} = 6.5310\) \(kg\cdot m/s\)  \\
\hline

4 & Planck force & \(F_{P} = \frac{E_{P}}{l_{P}} = 1.2103 \times 10^{44}\) \(N\) & \(F_{P} = E_{P}\cdot \sqrt{\frac{k_{\beta}m_{P}T_{P}}{\hbar^{2}}} = 1.2109 \times 10^{44}\) \(N\)  \\
\hline

5 & Planck energy & \(E_{P} = m_{P}c^{2} = 1.9561 \times 10^{9}\) \(J \) & \(E_{P} = \frac{\hbar^{2} c^{2}}{k_{\beta} T_{P}l_{P}^{2}} = 1.9593 \times 10^{9}\) \(J\) \\
\hline

6 & Planck frequency & \(f_{P} = \frac{c}{l_{P}} = 1.8549 \times 10^{43}\) \(Hz\) & \(f_{P} = \frac{c \sqrt{k_{\beta} T_{P}m_{P}}}{\hbar} = 1.8560 \times 10^{43}\) \(Hz\) \\
\hline

7 & Planck charge & \(Q_{p} = \sqrt{4 \pi \varepsilon_{0} \hbar c} = 1.8755 \times 10^{-18}\) \(C \) & \(Q_{P} = \left(4 \pi \varepsilon_{0} \sqrt{k_{\beta}T_{P} l_{P}^{2} m_{P}}      c\right)^{\frac{1}{2}} = 1.8745 \times 10^{-18}\) \(C\) \\
\hline
\end{tabular}
\end{center}
\begin{center}
Table 3: Derived Planck scale physical quantities.
\end{center}

\section*{New quantum spin perspective and Planck star}

The Planck star, proposed in the loop quantum gravity, is an important object that is resided at the singularity point of the black hole. The Planck star is created when the energy density becomes of the order of Planckian and at this point the quantum gravitational pressure prevents the gravitational collapse. This whole process takes many years for an outside observer; but, it takes few seconds in the local frame of reference. The Planck star is supposed be a solution of the black hole information paradox in the loop quantum gravity. Using the new quantum spin perspective, we propose new formulas regarding the theory of the Planck star [21].

The  actual formula of the size of the Planck star is written as

\begin{eqnarray}
r \sim \left(\frac{m}{m_{P}}\right)^{n} l_{P}
\end{eqnarray}

Where \(l_{P}\) is the Planck length, \(m\) is the initial mass for the Planck star, \(m_{P}\) is the Planck mass and \(n\) is the positive integer.  

From table (2), using the novel formula of the Planck length and the Planck mass, the formula of the size of the Planck star can be expressed as,

\begin{eqnarray}
r \sim \left(\frac{m k_{\beta} l_{P}^{2}T_{P}}{\hbar^{2}}\right)^{n} \sqrt{\frac{\hbar^{2}}{k_{\beta}T_{P}m_{P}}}
\end{eqnarray}

Since the curvature and the Planckian energy density are related to each other (\(R \sim 8 \pi \rho_{P}\)), it can also be expressed through novel mathematical expression (from table (3)), i.e., 

\begin{eqnarray}
R \sim 8 \pi m_{P}\cdot \left(\frac{k_{\beta} T_{P}m_{P}}{\hbar^{2}}\right)^{\frac{3}{2}}
\end{eqnarray}  

For \(n = \frac{1}{3}\), the size of the Planck star and the surface area of the Planck star are \(\left(\frac{m}{m_{P}}\right)^{\frac{1}{3}} l_{P}\) and \(A = \left(\frac{m}{m_{P}}\right)^{\frac{2}{3}} l_{P}^{2}\) respectively. Hence, the formula of  surface area of the Planck star using new quantum spin perspective takes the form,

\begin{eqnarray}
A = \left(\frac{m k_{\beta} l_{P}^{2}T_{P}}{\hbar^{2}}\right)^{\frac{2}{3}} \frac{\hbar^{2}}{k_{\beta}T_{P}m_{P}}
\end{eqnarray}

The life time of the black hole having mass of the order of \(10^{12} kg\), is \(t_{BH} = 14 \times 10^{9} years\). The size of of this black hole can be expressed as,

\begin{eqnarray}
r^{3} \sim \frac{G\hbar}{348 \pi c^{2}}t_{BH}
\end{eqnarray} 

Since, \(l_{P} = \frac{G\hbar}{c^{3}}\) and \(t_{P} = \frac{l_{P}}{c}\);

The equation becomes,

\begin{eqnarray}
r^{3} \sim \frac{l_{P}^{2}c}{348 \pi}t_{BH} = \frac{t_{BH} l_{P}^{3}}{348 \pi t_{P}}
\end{eqnarray} 

From table (2), considering the formula of the Planck length \(l_{P}\), the equation becomes, 

\begin{eqnarray}
r = \sqrt[3]{\frac{t_{BH}}{348 \pi t_{P}}}l_{P} = \sqrt[3]{\frac{t_{BH}}{348 \pi t_{P}}}\sqrt{\frac{\hbar^{2}}{k_{\beta} T_{P} m_{P}}}
\end{eqnarray}

From this modification, one can see that the size, the surface area involves some new Planckian quantities that adds new significance to this topic. Since, table (1) to table (3) are validated; this modification does not change the significance of the actual formula.

\section*{Other consequences of this novel quantum spin perspective in quantum gravity}

The quantum gravitational effect begins to appear when the Planck scale is considered. Therefore, this perspective is crucial at the quantum gravity scale; especially, for the physics of big bang and black hole.  At the Planck scale, for mutual small change, the Planck temperature \(T_{P}\), the Planck mass \(m_{P}\) and the Planck length \(l_{P}\) are related to the quantum spin.   Therefore, we establish relationship of the quantum spin with the Planck temperature \(T_{P}\), the Planck mass \(m_{P}\) and the Planck length \(l_{P}\). Here we take two Planckian quantities as constant; while establishing the relationship of the quantum spin with the third Planck quantity. For instance, the Planck mass and the Planck length are taken as constant; while establishing the relationship between the Planck temperature and the quantum spin. Here, it should be noted that the change in the value of the Planck temperature \(T_{P}\), the Planck mass \(m_{P}\) and the Planck length \(l_{P}\) with the integer \(n\) is very small at the Planck scale; but, it adds new notion for the Planckian physics i.e., quantum gravity. 

\section*{Planck temperature and quantum spin}

One can also validate the equation (7) by  calculating the value of \(T_{P}\), \(l_{P}\) and \(m_{P}\). For Instance, the value of the Planck temperature from the equation (7) can be calculated; i.e., 

\begin{eqnarray}
\therefore T_{P} = \frac{2^{2} \times \frac{n^{2}\hbar^{2}}{2^{2}}}{k_{\beta} l_{P}^{2} m_{P}}= 1.419 \times 10^{32} K \approx 1.416 \times 10^{32}  K
\end{eqnarray}

Hence, this value is almost equal to actual value of \(T_{P}\) [17-20]. In the above formula, \(\hbar, k_{\beta}, l_{P}^{2},\) and \(m_{P}\) are constant at the Planck scale. Hence, the Planck temperature depends on \(n^{2}\). therefore,

\begin{eqnarray}
\therefore T_{P} = \mathcal{R}_{1} n^{2} \rightarrow  T_{P} \propto n^{2}
\end{eqnarray}

Where \(\mathcal{R}_{1} = \frac{\hbar^{2}}{k_{\beta} l_{P}^{2} m_{P}}\) is a constant. The value of \(\mathcal{R}_{1}\) is \(1.419 \times 10^{32} K \). Thus, the value of Planck temperature \( T_{P}\)  changes with the square of integer \(n\) at the Planck scale. With the increment in the value of the integer \(n^{2}\), the value of the Planck temperature \(T_{P}\) also increases. The nature of such a plot \(T_{P} \rightarrow n^{2}\) is linear. This relationship is here given in the tabular form. 

\begin{center}
\begin{tabular}{{| p{0.04\textwidth} | p{0.05\textwidth}| p{0.2\textwidth} |}} 
\hline
\(n\) & \(n^{2}\) & \(T_{P} = \frac{\hbar^{2}}{k_{\beta} l_{P}^{2} m_{P}} n^{2} \)   \\
\hline\hline
1 & 1 & \(1.419 \times 10^{32} K\) \\
\hline
2 & 4 & \(5.676 \times 10^{32} K\) \\
\hline
3 & 9 & \(12.771 \times 10^{32} K\) \\
\hline
4 & 16 & \(22.704 \times 10^{32} K\) \\
\hline
\end{tabular}
\end{center}
\begin{center}
Table 4: The relationship between the Planck temperature \(T_{P}\) and the integer \(n\).
\end{center}

\begin{figure}
\begin{center}
\includegraphics[scale=0.5]{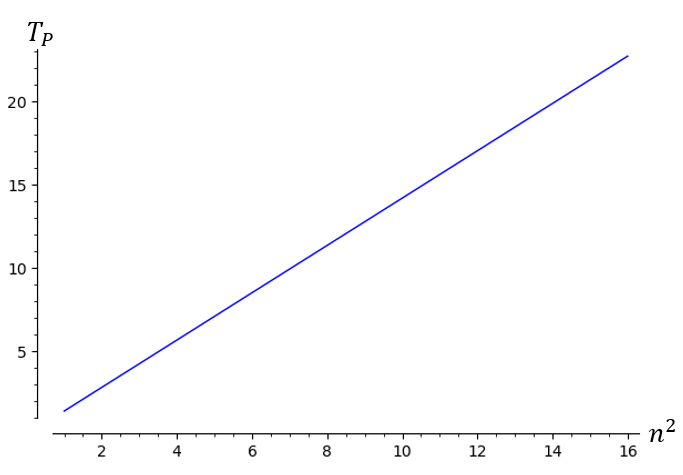}
\end{center}
\begin{center}
Fig. 1: The plot of \(T_{P} \rightarrow n^{2}\). 
\end{center}
\end{figure}

\bigskip
\bigskip
\bigskip
\bigskip
\bigskip

\section*{Planck mass and quantum spin}

Similar to \(T_{P}\), the value of \(m_{P}\) can be found using the equation (7), i.e., 

\begin{eqnarray}
m_{P} = \frac{2^{2} \times \frac{n^{2}\hbar^{2}}{2^{2}}}{k_{\beta} l_{P}^{2} T_{P}} = 2.181 \times 10^{-8} kg \approx 2.176 \times 10^{-8} kg
\end{eqnarray}

Hence, this value is almost equal to actual value of \(m_{P}\) [17-20]. \(\hbar\), \(k_{\beta}\), \(l_{P}^{2}\) and \(T_{P}\) are constant at the Planck scale. Thus, the Planck mass depends on \(n^{2}\). Therefore, 

\begin{eqnarray}
\therefore m_{P} = \mathcal{R}_{2} n^{2} \rightarrow m_{P} \propto n^{2} 
\end{eqnarray}

Where, \(\mathcal{R}_{2} = \frac{\hbar^{2}}{k_{\beta} l_{P}^{2} T_{P}}\) is a constant. The value of \(\mathcal{R}_{2}\) is \(2.181 \times 10^{-8} kg\). Thus, the value of the Planck mass changes with the square of integer \(n\) at the Planck scale. With the increment in the value of the integer \(n^{2}\), the value of the Planck temperature \(m_{P}\) also increases. The nature of such a plot \(m_{P} \rightarrow n^{2}\) is linear. This relationship is here given in the tabular form. 

\begin{center}
\begin{tabular}{{| p{0.04\textwidth} | p{0.05\textwidth}| p{0.2\textwidth} |}} 
\hline
\(n\) & \(n^{2}\) & \(m_{P} = \frac{\hbar^{2}}{k_{\beta} l_{P}^{2} T_{P}} n^{2} \)   \\
\hline\hline
1 & 1 & \( 2.176 \times 10^{-8} kg \) \\
\hline
2 & 4 & \( 8.724 \times 10^{-8} kg \) \\
\hline
3 & 9 & \( 19.629 \times 10^{-8} kg \) \\
\hline
4 & 16 & \(34.896 \times 10^{-8} kg\) \\
\hline
\end{tabular}
\end{center}
\begin{center}
Table 5: The relationship between the Planck temperature \(m_{P}\) and the integer \(n\).
\end{center}

\begin{figure}
\begin{center}
\includegraphics[scale=0.5]{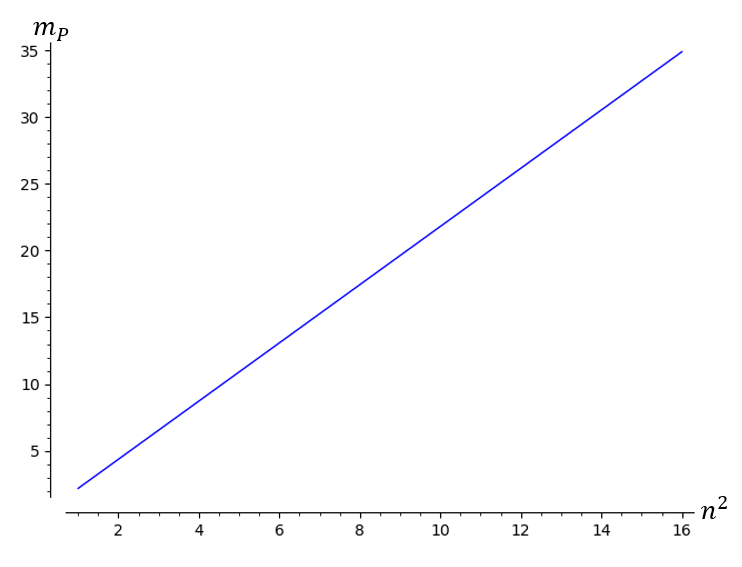}
\end{center}
\begin{center}
Fig. 2: The plot of \(m_{P} \rightarrow n^{2}\). 
\end{center}
\end{figure}

\bigskip
\bigskip
\bigskip
\bigskip
\bigskip

\section*{Planck length and quantum spin}

Similar to \(T_{P}\), the value of \(l_{P}\) can be found using the equation (7), i.e.,

\begin{eqnarray}
l_{P} = \sqrt{\frac{2^{2} \times \frac{n^{2}\hbar^{2}}{2^{2}}}{k_{\beta} T_{P} m_{P}}} =  1.618 \times 10^{-35} m \approx 1.616 \times 10^{-35} m
\end{eqnarray} 

Hence, this value is almost equal to actual value of \(l_{P}\) [17-20]. Here, \(\hbar\), \(k_{\beta}\), \(m_{P}\) and \(T_{P}\) are constant at the Planck scale. Thus, the Planck length depends on \(n\). therefore, 

\begin{eqnarray}
\therefore l_{P} = \mathcal{R}_{3} n \rightarrow l_{P} \propto n
\end{eqnarray}

Where, \(\mathcal{R}_{3} = \sqrt{\frac{\hbar^{2}}{k_{\beta} T_{P} m_{P}}}\) is a constant.  The value of \(\mathcal{R}_{3}\) is \(1.618 \times 10^{-35} m \). Thus, the value of \(l_{P}\) changes with the integer \(n\) at the Planck scale. With the increment in the value of the integer \(n\), the value of the Planck temperature \(l_{P}\) also increases. The nature of such a plot \(l_{P} \rightarrow n\) is linear. This relationship is here given in the tabular form. 

\begin{center}
\begin{tabular}{{| p{0.05\textwidth}| p{0.2\textwidth} |}} 
\hline
\(n\) & \(l_{P} = \sqrt{\frac{\hbar^{2}}{k_{\beta} T_{P} m_{P}}} n \)   \\
\hline\hline
1 & \( 1.618 \times 10^{-35} m \) \\
\hline
2 & \( 3.236 \times 10^{-35} m \) \\
\hline
3 & \( 4.854 \times 10^{-35} m \) \\
\hline
4 & \(6.472 \times 10^{-35} m\) \\
\hline
\end{tabular}
\end{center}
\begin{center}
Table 6: The relationship between the Planck temperature \(l_{P}\) and the integer \(n\).
\end{center}

\begin{figure}
\begin{center}
\includegraphics[scale=0.5]{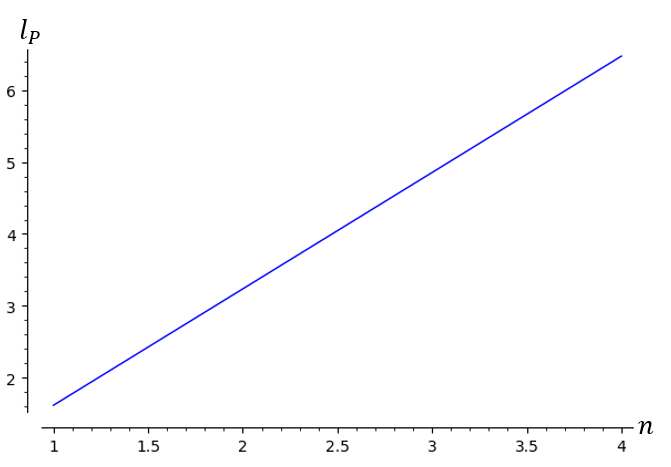}
\end{center}
\begin{center}
Fig. 3: The plot of \(l_{P} \rightarrow n\).. 
\end{center}
\end{figure}

\bigskip
\bigskip
\bigskip
\bigskip
\bigskip

\section*{Conclusions}

This novel perspective of the quantum spin has far reaching consequences in the theory of quantum gravity. Consequences of this novel perspective in LQG is known. With the aid of this novel perspective, the new formulas are derived for the reduced Planck constant, the Boltzmann constant \(k_{\beta}\), the gravitation constant \(G\), the fine structure constant \(\alpha\) and the speed of light \(c\). Moreover, the value of various these universal constants can be validated by comparing the derived formula from the quantum spin with the old formula. we also find new formulas for fundamental Planckian quantities and the derived Planckian quantities using this novel perspective. Novel formulas for the Planck star such as the size, the curvature, the surface area and the size of black hole (for the Planck star) are proposed  without modifying its significance.  From this perspective, we establish relationship of the quantum spin \(i.e., n\) with the Planck temperature \(T_{P}\), the Planck mass \(m_{P}\), and the Planck length \(l_{P}\) at the Planck scale. The plots such as \(T_{P} \rightarrow n^{2}\), \(m_{P} \rightarrow n^{2}\) and \(l_{P} \rightarrow n\) are studied that add the new notion to in the physics of quantum gravity . Therefore, this perspective shows the importance of the quantum spin at the quantum gravity scale.  

\section*{Acknowledgment}

The authors are thankful to Physics Department, Saurashtra University, Rajkot, India.

\end{document}